\documentclass[acmtog, 10pt]{acmart}
\acmSubmissionID{461}

\usepackage{booktabs} 

\usepackage{wrapfig}

\usepackage[ruled]{algorithm2e} 
\usepackage{amsmath,bm} 

\SetAlFnt{\small}
\SetAlCapFnt{\small}
\SetAlCapNameFnt{\small}
\SetAlCapHSkip{0pt}

\acmJournal{TOG}

\newcommand{\Comment}[1]{}

\setcopyright{none}
\renewcommand\footnotetextcopyrightpermission[1]{}
\begin{document}
\title{Survey: Machine Learning in Production Rendering}

\author{Shilin Zhu}
\affiliation{
\institution{University of California San Diego}}
\email{shz338@eng.ucsd.edu, https://cseweb.ucsd.edu/~shz338/}



\begin{abstract}
\vspace{+3mm}
\normalsize
  In the past few years, machine learning-based approaches have had some great success for rendering animated feature films. This survey summarizes several of the most dramatic improvements in using deep neural networks over traditional rendering methods, such as better image quality and lower computational overhead. More specifically, this survey covers the fundamental principles of machine learning and its applications, such as denoising, path guiding, rendering participating media, and other notoriously difficult light transport situations. Some of these techniques have already been used in the latest released animations while others are still in the continuing development by researchers in both academia and movie studios. Although learning-based rendering methods still have some open issues, they have already demonstrated promising performance in multiple parts of the rendering pipeline, and people are continuously making new attempts. 
\end{abstract}

\maketitle

\textbf{Author}

Shilin Zhu is a Ph.D. student at the University of California San Diego. He works on deep learning for computer graphics and 3D computer vision.

\section{Introduction}
\label{sec:intro}
Physically-based rendering \cite{pharr2016physically} is crucial for generating artistic and photorealistic imagery in feature animated films, thanks to the capability of computing costly global illumination and complex light transport by modern rendering algorithms such as Monte Carlo (MC) path tracing \cite{kajiya1986rendering,veach1997robust, fascione2017path, fascione2017path2, fascione2019path} and particle density based photon mapping (PM) \cite{jensen1996global, hachisuka2009stochastic}. 

Despite the beautiful look of final rendered images, the movie production industry is complaining about the high complexity of light transport simulation algorithms. In general, both MC based and particle-based approaches require simulating millions or even billions of light paths (or photons) to produce noise-free images, which are too computationally expensive to generate regular 90/120-minute animations. For example, it takes hundreds of CPU hours to render a single frame with stunning volumetric effects in Frozen 2 (\copyright 2019 Disney). Such expensive computational overhead forces movie studios to keep giant rendering farms with a massive amount of CPU cores as well as re-organize the code to achieve the maximum parallelism. In fact, light transport simulations usually involve lots of redundant computation. These brute-force light transport algorithms significantly slow down the production speed of artists. Therefore, movie studios have done many engineering tricks to follow the continuing increase of the scene complexity for every new show \cite{burley2018design, christensen2018renderman, fascione2017path, fascione2017path2, fascione2019path}.

Recently, some significant improvements have been achieved by introducing machine learning into the rendering pipeline \cite{keller2018machine}. Neural networks are good at learning complicated functions such as the distribution of light in the scene, or learning to clean up the residual noise in the rendered image, and can be executed very fast on GPUs. These advantages make deep learning a very competitive approach for highly complex light transport conditions, which could otherwise take much longer time to render if using traditional methods to achieve comparable image quality. 

This paper covers several most significant improvements in deep learning assisted rendering in recent years. As mentioned, traditional MC path tracing requires hundreds or even thousands of samples per pixel to effectively reduce the variance of the estimated pixel color (i.e., MC noise \cite{kajiya1986rendering, lafortune1996mathematical}) to an acceptable level. However, the variance reduction speed gets slower and slower with the increasing number of samples \cite{mitchell1996consequences}. Such diminishing returns lead to a huge amount of samples required to make the final rendered image noise-free to human eyes. Recently, people have tried using Convolutional Neural Networks (CNN) \cite{lecun1989backpropagation, lecun1998gradient} to denoise the images with MC artifacts. The neural net takes the low-sample-count noisy image with some additional geometric and material parameters that can be quickly rendered and outputs the clean image without any visible artifact. This image-space technique has been adopted by various studios and used in most latest animations \cite{bako2017kernel, vogels2018denoising}.

Although denoising significantly reduces the overall rendering time, it still takes a while to generate an initial noisy image with all the details presented. When details are missing due to lack of samples, we also cannot rely on neural network denoiser to generate contents that were not initially intended by artists. Furthermore, in some conditions, the scene can be too complex even to render a reasonable low-sample-count image. For example, when light sources only illuminate the scene indirectly, it is challenging to construct light paths to the camera by standard path tracing (e.g., a light source behind a slightly opened door). The problem could be resolved by generating better samples that could always form paths toward light sources. To achieve this goal, people have proposed the path guiding technique that lets a simple neural net to accurately estimate the local radiance distribution, which then is used for sampling better light paths \cite{vorba2019path}. Path guiding has been proved to use many fewer samples to achieve high rendering quality, and it has been deployed in Frozen 2 (\copyright 2019 Disney) \cite{muller2017practical}.

Besides denoising and path guiding, there are some exceptional severe light transport cases where traditional rendering algorithms are inefficient, but machine learning makes those tasks a lot easier. As for movie production, water and volumetric effects are the hardest parts to render \cite{burley2018design}, since light can be scattered many times before it can reach the camera. Researchers designed a unique deep learning framework to estimate the multi-scattering radiance contribution from indirect illumination, leading to significant speedup for rendering high-albedo clouds and thin smokes \cite{kallweit2017deep}. Additionally, there are special caustics effects in water and volume that cannot be generated efficiently with path tracing. Instead, people add an extra photon mapping pass \cite{jensen1996global, hachisuka2009stochastic} to render caustics and compose it with path tracing results \cite{christensen2016path}. Recently some works use 3D deep learning techniques to do particle density estimation in photon mapping \cite{shilin2020deep} and works that guide photon emission in order to create caustics much faster \cite{peter1998importance, vorba2019path}. Despite that neural networks could obtain good performance in these special cases, there are still some subtle artifacts across pixels when machine learning methods are enabled. Therefore, the industry is still working on further improving them.

There is certainly more work than what this survey can cover in detail, such as machine learning for gradient-domain rendering \cite{guo2019gradnet} and generating high-quality high-dimensional random point samples \cite{leimkuhler2019deep}. The rest of the paper is organized as follows. Section 2 introduces the basic principles of machine learning, deep neural networks, and production rendering. Section 3 and 4 provide a detailed description of denoising and path guiding techniques published in the past few years. In section 5, some other developed techniques which could be used by studios in the future are discussed. Section 6 concludes the survey.

\vspace{-3mm}
\section{Background}
\label{sec:basic}

\subsection{Fundamentals of machine learning}
In general, machine learning (ML) takes some statistical data and extracts useful patterns or features out of it, based on the objective that is given, through iterative optimization. In the early years of machine learning development, people needed to design hand-crafted features, and ML models learned feature parameters (e.g., parametric mixture models \cite{reynolds2009gaussian}). Although this direction of work achieved many great improvements, people then found that using such hardcoding features injects too much human knowledge, which limits the capability of ML systems. Neural networks were then developed so that people do not have to control the pattern. Instead, the learning system itself was responsible for handling it. It has been proved that neural networks are universal function approximators $\mathcal{F}$ \cite{leshno1993multilayer}, mapping the input $x$ to the estimated output $\hat{y}$ with learned parameters $\theta$:
\begin{align}
    \hat{y} = \mathcal{F}(x; \theta)
    \label{eqn:nn}
\end{align}
If we provide ground-truth value $y$ and design a loss function $\mathcal{L}(\hat{y}, y)$ (e.g., mean-squared error (MSE) or mean absolute error (MAE)), the neural network will adjust its parameters to minimize the difference between the predicted output and ground-truth through gradient-descent optimization \cite{ruder2016overview, goodfellow2016deep}. In rendering, the input-output $(x, y)$ pair could be generated by the renderer, and ML learns a nonlinear mapping function between them. For example, the input could be low-sample-count noisy images, and output could be high-sample-count noise-free images. Among the huge amount of network architectures proposed by machine learning community, Multi-Layer Perceptron (MLP) \cite{rumelhart1985learning, rumelhart1986learning}, Convolutional Neural Net (CNN) \cite{lecun1989backpropagation, lecun1998gradient}, Autoencoder (AE) \cite{rumelhart1986learning, vincent2008extracting}, and Recurrent Neural Net (RNN) \cite{rumelhart1986learning, hochreiter1997long} are the most popular choices in computer graphics. MLP is often used to process unstructured data such as point clouds, while CNN operates on structured grid data like images. AE learns efficient and compact data encoding in an unsupervised manner, and such representation could be used for data compression or generative models. By adding internal states and connections along temporal sequences, RNNs can model the dynamics of sequential data and combine features from all time steps. It can learn temporal relations from a sequence of data and thus able to generate temporally-stable outputs (e.g., animation sequences without flickering).

However, sometimes it is impossible to obtain the ground-truth $y$ to supervise the ML system, in which case reinforcement learning (RL) \cite{sutton2018reinforcement} is usually considered. RL introduces the reward mechanism which is similar to how human learns to interact with the environment through trial and error. The RL system learns to figure out what next action to take based on the current state and expected rewards, and Q-learning is one of the most commonly used methods \cite{watkins1992q}. In fact, RL and Q-learning have deep mathematical relations with rendering, since they share the same integral equation that every rendering algorithm is trying to solve. 

Lecun et al. \shortcite{lecun2015deep} and Goodfellow et al. \shortcite{goodfellow2016deep} have given detailed and complete reviews of machine learning and deep neural networks. 

\Comment{
\begin{figure}[t]
    \includegraphics[width=\linewidth]{figs/denoise_0.pdf}
    \caption{Results of applying Disney's denoiser and Pixar's denoiser on Moana (\copyright 2016 Disney) and Finding Dory (\copyright 2016 Disney/Pixar). The raw images were rendered with 128 samples per pixel (SPP). These images were originally published in \cite{burley2018design} and \cite{christensen2018renderman}. }
    \label{fig:denoise_0}
\end{figure}
}

\subsection{Fundamentals of rendering}

\subsubsection{The rendering equation}
Basically, all the rendering algorithms aim to approximate the solution of the rendering equation (or Light Transport Equation, LTE) \cite{kajiya1986rendering, pharr2016physically}:
\begin{align}
    L(p, \omega_{o}) = L_{e}(p, \omega_{o}) + \int_{H}f(p, \omega_{o}, \omega_{i})L(\omega_{i})|cos(\theta_{i})|d\omega_{i}
    \label{eqn:re}
\end{align}
where $L(p, \omega_{o})$ is the outgoing radiance from surface shading point $p$ in direction $\omega_{o}$, $L_{e}(p, \omega_{o})$ is the emitted radiance at that point, $f(p, \omega_{o}, \omega_{i})$ is the Bidirectional Scattering Distribution Function (BSDF) \cite{bartell1981theory} that describes the way in which the light is scattered by the surface with specific materials, $L(\omega_{i})$ is the incoming radiance to point $p$ from direction $\omega_{i}$, and $\theta_{i}$ is the angle between ray direction and surface normal direction. The rendering equation describes how much radiance is redirected to the outgoing direction by a surface from the received incoming radiance. This is the solid angle form of LTE. We can also convert it to the surface area form by changing the integrated variable from $d\omega$ to $dA$:
\begin{align}
\begin{split}
    L(p_{i} \rightarrow p_{i-1}) = &L_{e}(p_{i} \rightarrow p_{i-1}) + \int_{A}f(p_{i+1} \rightarrow p_{i} \rightarrow p_{i-1})\\&L(p_{i+1} \rightarrow p_{i})G(p_{i+1} \leftrightarrow p_{i})dA(p_{i+1})
    \label{eqn:re_a}
\end{split}
\end{align}
where $p_{i}$ is the $i$-th vertex (i.e., shading point) along the path. $G$ is a geometry term representing the relative orientation and visibility between two vertices.

\subsubsection{Monte-Carlo estimation}
For complex integrals such as the rendering equation (Equ.\ref{eqn:re}), often they do not have analytical solutions. Monte-Carlo (MC) method was invented to compute it numerically \cite{mackay1998introduction}:
\begin{align}
    L = \int f(x)dx \rightarrow F = \frac{1}{N}\sum_{i=1}^{N} \frac{f(x_{i})}{p(x_{i})}
    \label{eqn:mc_int}
\end{align}
where $F$ is the MC estimated result, $N$ is the total number of samples, and $p(x_{i})$ is the probability density of sampling $x_{i}$. Moreover, MC estimation is unbiased (i.e., $E[F] = L$) and consistent (i.e., $\lim_{N\rightarrow\infty}P(|F-L| > \epsilon) = 0$), which makes it an excellent estimator to approximate the true integration result. In MC theory, if the user-defined sampling distribution $p$ is proportional to the function $f$, we could get more samples from high-value regions of $f$. This is called importance sampling that significantly speeds up the convergence of MC estimation \cite{veach1995optimally, veach1997robust, kahn1955use}.

\subsubsection{REYES}
\label{sec:reyes}
Historically, physically-based path tracing was not used in movie production due to the lack of ability to handle complex geometries, achieve coherent memory access, and low-cost shading calculation. At that time, the most popular algorithm was REYES, where each surface was tessellated into pixel-sized micropolygons, and shading was performed on each polygon vertex \cite{burley2018design, christensen2018renderman, christensen2016path}. REYES was able to achieve great data coherency and shading reuse. The point-based method was also invented to approximate global illumination with additional data storage cost of precomputed point clouds \cite{christensen2010point}. 

\Comment{
\begin{figure}[t]
    \includegraphics[width=\linewidth]{figs/pg_illu.pdf}
    \caption{Path guiding predicts the incident light distribution to shoot more rays towards the light source where BSDF sampling sometimes could be inefficient. This image is modified from the original slides in \cite{vorba2019path}.}
    \label{fig:pg_illu}
\end{figure}

\begin{figure}[t]
    \includegraphics[width=\linewidth]{figs/pg.png}
    \caption{Path guiding rendering results for Alita: Battle Angel \copyright 2018 20th Century Fox. The caustics in Alita's eyes and on the lake bed of vast underwater scenes were difficult to render with path tracing. These images were originally published in \cite{vorba2019path}.}
    \label{fig:pg_0}
    \vspace{-4mm}
\end{figure}
}

\vspace{-2mm}
\subsubsection{Monte-Carlo path tracing}
\label{sec:mcpt}
Although REYES and point-based methods worked fine at that time for rendering movies like Toy Story 1 (1995 $\copyright$ Disney/Pixar) and Up (2009 $\copyright$ Disney/Pixar), they required multiple rendering passes where artists struggled with managing data like shadow maps and reflection maps. Furthermore, they were not well suited for progressive rendering and computing global illumination in a unified way. Therefore, almost all the studios switched to Monte-Carlo (MC) path tracing, since it only required a single pass to render all the effects with the quality continuously improving over time \cite{fascione2017path, fascione2017path2}. 

MC path tracing begins by shooting rays (i.e., image-space samples) from the camera origin. The rays are randomly sampled so that each one has a different direction. This sampling is usually done by generating low-discrepancy samples or other advanced sequences with excellent statistical properties (e.g., \cite{sobol1967distribution, christensen2018progressive}). Then the ray is traced into the scene and when it hits a surface, an intersection point is recorded. The local surface material (i.e., BSDF) and texture at that intersected shading point are evaluated. According to the sampling strategy (importance sampling or multiple importance sampling \cite{veach1995optimally, veach1997robust}), one or more new rays are spawned from the shading point and traced in the same manner, until the rays reach a light source or leave the scene permanently after multiple bounces. Then the complete paths between light sources and the camera are formed. Finally, the radiance contribution along each path is computed and combined using MC estimation (Equ.\ref{eqn:mc_int}) \cite{mackay1998introduction}, which is then converted to the pixel value. In summary, rays are generated from the camera and traced recursively in the scene. When the ray hits any light source, the radiance that contributes to the pixel through the path is evaluated.

 The rendering equation usually does not have an analytical solution. Hence, path tracing solves it numerically: The radiance contribution $L$ from a light source to the camera through a traced path is approximated by MC integration \cite{pharr2016physically}:
\begin{align}
\begin{split}
    L = &\frac{L_{e}(p_{l} \rightarrow p_{l-1})f(p_{l} \rightarrow p_{l-1} \rightarrow p_{l-2})G(p_{l} \leftrightarrow p_{l-1})}{p_{A}(p_{l})}\\
    & \times \prod_{j=1}^{l-2}\frac{f(p_{j+1} \rightarrow p_{j} \rightarrow p_{j-1})|cos(\theta_{j})|}{p_{\omega}(p_{j+1} \rightarrow p_{j})}
    \label{eqn:pt}
\end{split}
\end{align}
where $p_{l}$ is the sampled point on the light source, $L_{e}(p_{l} \rightarrow p_{l-1})$ is the radiance leaving the light source, $p_{\omega}$ represents the probability density of sampling outgoing ray directions for intermediate path vertices, and $p_{A}$ is the density of sampling a point over the surface of the light source. MC path tracing is an unbiased and consistent algorithm, since the expected radiance estimation $L$ is equal to the real solution of LTE (Equ.\ref{eqn:re}) regardless of how many samples/rays are used, and the variance of $L$ decreases with number of samples/rays. However, the brute-force path tracing converges slowly and usually requires tracing thousands of samples per pixel (SPP). Otherwise, the result images will contain MC noise.

Ideally, if we perform importance sampling (Equ.\ref{eqn:mc_int}) and the probability density is proportional to the integrand, then MC estimation only need one sample to compute the exact solution with zero variance \cite{veach1997robust}:
\begin{align}
    p_{\omega} \propto f(p, \omega_{o}, \omega_{i})L(\omega_{i})|cos(\theta_{i})|
    \label{eqn:mc}
\end{align}
However, in most cases, we do not know the incoming radiance $L(\omega_{i})$ before we continue tracing more rays. Thus path tracing usually does BSDF importance sampling as 
\begin{align}
    p_{\omega} \propto f(p, \omega_{o}, \omega_{i})|cos(\theta_{i})|
    \label{eqn:is_brdf}
\end{align}
and combines it with only direct light sampling (i.e., next event estimation \cite{veach1995optimally, pharr2016physically}).

\Comment{
\begin{figure}[t]
    \includegraphics[width=\linewidth]{figs/caustics_0.pdf}
    \caption{Examples of scenes with caustics effects. Moana \copyright 2016 Disney. A small rock under the water rendered using guided photon mapping (this image is originally published in \cite{burley2018design}).}
    \label{fig:caustics_0}
    \vspace{-3mm}
\end{figure}
}

\subsubsection{Path guiding}
\label{sec:pg_basic}
Path tracing is not a very efficient algorithm: Lots of samples are wasted due to zero radiance contribution to the pixels when those rays leave the scene without hitting any light source. We do not know in advance which ray direction could connect to the light source, so path tracing sampling is sub-optimal (Equ.\ref{eqn:is_brdf}), even with combined direct light sampling. In some difficult light transport scenarios, path tracing could completely fail to find the indirect light source. To improve the sampling efficiency, path guiding was proposed to predict where the light comes from at each point. Path guiding \cite{vorba2019path} starts with uniform sampling and progressively learns the incident radiance distribution at each shading point from recorded historical samples, and uses the learned sampling distribution to guide light transport directions toward light sources:
\begin{align}
    p_{\omega} \propto L(\omega_{i})
    \label{eqn:pg}
\end{align}
Since our goal is to achieve Equ.\ref{eqn:mc}, we combine two importance sampling strategies (Equ.\ref{eqn:is_brdf} and Equ.\ref{eqn:pg}) using multiple importance sampling (MIS) \cite{veach1997robust}. The incident radiance distribution could be represented by mixture models or any learned probability density function. People have tried both online / reinforcement \cite{muller2017practical} and offline learning \cite{bako2019offline} to produce better sampling distributions, and both have achieved improved rendering quality with many fewer samples/rays compared with vanilla path tracing. Frozen 2 (\copyright 2019 Disney) was the very first animated feature film that adopted this technique.

\subsubsection{Photon mapping}
\label{sec:pm_basic}
Although path tracing is simple and elegant, it is very hard or sometimes even impossible to effectively render some special light transport effects, such as caustics. The light could be transmitted through transparent objects like water in the swimming pool and form bright spots on a diffuse surface underneath. These paths involve multiple specular reflections/refractions and can have a very low probability of being sampled using MC path tracing, but they are crucial for creating realistic imagery. In the extreme case, when there is only one path that can reach the light source after multiple specular bounces, it is impossible for MC path tracing to sample it (i.e., delta function). In some movies such as Alita: Battle Angel (\copyright 2018 Twentieth Century Fox) and Moana (\copyright 2016 Disney), some scenes have water, so caustics must be generated efficiently. To achieve the goal, people have developed Photon Mapping (PM) \cite{jensen1996global} and Stochastic Progressive Photon Mapping (SPPM) \cite{hachisuka2009stochastic} methods. First, photons are emitted from light sources and stored on scene surfaces in the pre-pass. In the second pass, rays are traced from the camera and radiance is estimated at each shading point by summing up energies of its nearby photons:
\vspace{-3.1mm}
\begin{align}
    L(p, \omega_{o}) \approx \frac{1}{N} \sum_{i=1}^{N} k(p, p_t) \tau_t,
    \label{eqn:pm}
\end{align}
where $N$ is the total number of photon paths that are emitted in a scene, 
$p_t$ is the location of a photon, $\tau_t$ is the photon contribution and $k$ represents a kernel function. In general, the photon contribution $\tau_t$ is the product of the BSDF and the photon energy. PM is especially effective in generating caustics since there are a huge number of photons stored in those areas. This makes it efficient to create caustics when the light gets reflected and refracted specularly. In movie production, the common way is to compose path tracing results and photon mapping caustics after they are separately rendered \cite{christensen2016path}. Some studios are working on more advanced topics such as guiding photon emissions so that photons will have high probability of landing in the area that is visible to the camera \cite{vorba2019path}. 

There are many more topics on production rendering such as ray batch and sorting, memory footprint, efficient light sampling, handling massive geometry, texture management, volume, subsurface scattering, hair/fur, etc. Please refer to \cite{burley2018design, christensen2018renderman, fascione2017path, fascione2017path2, fascione2019path} for a complete description. 

\section{Denoising Monte-Carlo Images}
\label{sec:denoising}
\Comment{
\begin{figure*}[h]
    \includegraphics[width=\textwidth]{figs/denoiser_1.pdf}
    \caption{Denoising results of Disney and Pixar's kernel-predicting neural networks. The raw images were rendered with 16/32 SPP. Finding Dory \copyright 2016 Disney/Pixar. Cars 3 \copyright 2017 Disney/Pixar. Coco \copyright 2017 Disney/Pixar. Olaf's Frozen Adventure \copyright 2017 Disney. These images were originally published in \cite{bako2017kernel} and  \cite{vogels2018denoising}.}
    \label{fig:denoise_1}
    \vspace{-2mm}
\end{figure*}
}

As mentioned earlier in Sec.\ref{sec:mcpt}, path tracing rendered images often have remaining MC noise due to insufficient samples/rays per pixel. The estimation variance is reduced slower with more samples (diminishing returns \cite{mitchell1996consequences}). Image-space and sample-space denoising become the most effective solutions to deal with MC noise, and they are easily integrated into the existing production pipeline. 

\vspace{-2mm}
\subsection{Traditional denoiser}
Denoising MC renderings can be traced back to Rushmeier et al. \shortcite{rushmeier1994energy} using hand-designed nonlinear filters. The idea of adding auxiliary features such as normal and depth maps to improve the denoising performance was originally proposed by \cite{mccool1999anisotropic}. Later kernel-weighted filters with or without additional features were developed, and most of them were based on the joint cross-bilateral and/or non-local means approach \cite{sen2012filtering, moon2013robust, rousselle2013robust, buades2008nonlocal, rousselle2012adaptive, buades2005review, zimmer2015path}. Some first-order and higher-order regression methods were also explored in the context of denoising MC images \cite{bitterli2016nonlinearly, moon2014adaptive, moon2016adaptive, bauszat2011guided}. Also, there is work on sheared filtering based on light transport frequency analysis that can determine near-optimal sampling rates for different rendering effects \cite{yan2015fast, egan2009frequency, wu2017multiple}. For real-time interactive rendering denoising, people have tried temporal sample accumulation and reuse to increase the equivalent sample counts using a sequence of frames \cite{schied2017spatiotemporal, schied2018gradient}.

Adaptive sampling and reconstruction distributes the samples according to local frequencies and contrasts, where more samples were assigned to high-frequency or high-contrast regions to capture details and reduce overall noise \cite{hachisuka2008multidimensional, overbeck2009adaptive}. In other words, we can redistribute samples and donate more to problematic pixels where light transport is hard. The adaptive sampling concept was then extended to different types of filters by selecting more effective kernels based on noise or variance estimation \cite{rousselle2011adaptive, moon2014adaptive, li2012sure, bitterli2016nonlinearly, rousselle2012adaptive, kalantari2013removing}. Zwicker et al. \shortcite{zwicker2015recent} provides a complete review of the adaptive sampling scheme.

\vspace{-2mm}
\subsection{Learning-based denoiser}
Despite the superior denoising performance of traditional kernel-based filters, they still cannot preserve the detailed contents well enough, and sometimes they over-blurred the entire image. In the past few years, many machine learning systems have been proposed to overcome this problem, thanks to the rapid development of deep neural networks.

Kalantari et al. \shortcite{kalantari2015machine} was an early work on bridging neural networks with traditional kernel-based filters. They used a simple but effective MLP network to drive the parameters of cross-bilateral and non-local means filters. The model was trained offline by pairs of noisy MC images and reference noise-free images rendered from a small set of scenes. It could then be used to filter an image of a new scene in only a few seconds. The major limitation of this work was the flexibility of their system since the type of filter was hardcoded.

The development of CNNs \cite{lecun1989backpropagation, lecun1998gradient} has made great success in natural image denoising and image super-resolution. Disney and Pixar \cite{bako2017kernel} proposed the first pioneering work of applying CNN to MC denoising. In this work, they focused on handling several unique problems when CNN was used. First of all, the network will also over-blur the image if it cannot distinguish well between scene content and MC noise. The paper resolved this issue by adding multiple auxiliary features (normal, depth, albedo, and their corresponding variances) and a novel diffuse/specular decomposition framework. In addition, their network directly learned the irradiance map, which was then multiplied by the diffuse albedo. This design preserved the scene details since it prevented the network from touching the high-frequency textures. Second, MC renderings are High Dynamic Range (HDR) images, which could affect the stability of neural network training and lead to color artifacts after denoising. To solve this problem, they proposed a novel kernel prediction architecture with proper normalization to keep the training stable. Their kernel predicting CNN was trained on 600 selected representative frames from the movie Finding Dory (\copyright 2016 Disney/Pixar) and tested on 25 frames from Cars 3 (\copyright 2017 Disney/Pixar) and Coco (\copyright 2017 Disney/Pixar). Later these studios extended the work to temporal denoising and added several new modules \cite{vogels2018denoising}. In this work, they added residual links and multi-scale spatial-temporal feature extractors to the network architecture, which enabled the new denoising CNN to remove low-frequency artifacts effectively. Moreover, the training was performed with an asymmetric loss function that not only helped to preserve detailed contents but also allowed artists to control the denoising strength through bias-variance trade-offs. To deal with the remaining artifacts after denoising, they also trained an extra small noise prediction network on the denoising results and used its predicted error map to drive adaptive sampling. The complete network was trained on Moana (\copyright 2016 Disney) and Cars 3 (\copyright 2017 Disney/Pixar), and tested on Olaf (\copyright 2017 Disney) and Coco (\copyright 2017 Disney/Pixar). These denoising networks are actively being used and further developed for future Disney and Pixar shows.

Besides studio research, there were also works in academia and other institutions for denoising and reconstructing MC renderings. Chaitanya et al. \shortcite{chaitanya2017interactive} proposed a recurrent convolutional architecture that could enforce temporal stability through mixing information from a sequence of frames. They used a mixed loss of the pixel color, gradient, and temporal consistency to train the network that preserved edges and avoided flickering. The network was tested on very low-sample-count images (4 SPP) and could denoise with interactive rates. This work could benefit both the real-time gaming industry and interactive preview for artists in the movie industry. Xu et al. \shortcite{xu2019adversarial} further added Generative Adversarial Networks (GAN) for training denoisers. They showed that GAN could help neural networks to generate noise-free images with more realistic high-frequency details and global illumination effects. Kuznetsov et al. \shortcite{kuznetsov2018deep} predicted adaptive sampling maps and reconstructed MC renderings using two separate CNNs, and produced high-quality results with extremely low sample counts. To train the entire system end-to-end, they presented a novel method to compute gradients with respect to adaptive sampling parameters (e.g., sample counts). Besides screen-space methods, sample-space denoising was also investigated. Gharbi et al. \shortcite{gharbi2019sample} focused on individual samples and used a combined MLP and CNN network to predict a kernel that splatted the radiance contribution of each sample onto the neighboring image pixels. Denoising in the sample space could make better use of individual sample information, compared with screen-space methods which mainly used only the statistics of groups of samples. 

Training these neural nets requires a large and diverse dataset since they were supervised learning. However, rendering high-sample-count ground-truth reference images was time-consuming and prevented the dataset from scaling up. Fortunately, recent works showed that training denoisers could be done using low-sample-count noisy images alone without requiring clean reference images \cite{lehtinen2018noise2noise, krull2019noise2void}, which made it possible to train the denoiser faster than before with an easier-generated dataset. 

There are still open issues when denoisers are applied in studio production \cite{burley2018design, christensen2018renderman}. For some movies like Zootopia (\copyright 2016 Disney) involving lots of hair and fur, the denoiser did not perform well since it never saw such data before, and some features such as normal maps were problematic. Similar problems were observed on shiny micro-bumped surfaces in Toy Story 4 (\copyright 2019 Disney/Pixar), where the denoiser falsely removed out those small shiny details. The current straightforward solution is to retrain the whole ML system for each of these failure cases. However, there should be a better way to improve the generalizability of pre-trained denoisers. Furthermore, with the recent boom of 3D deep learning \cite{chang2015shapenet, qi2017pointnet}, it is no doubt that denoising in 3D space directly could be another effective way.

\vspace{-1mm}
\section{Path guiding}
\label{sec:pg}

\Comment{
\begin{figure}[!t]
    \includegraphics[width=\linewidth]{figs/pg_0.pdf}
    \caption{Example production frame rendered using "practical path guiding" (PPG) algorithm \cite{muller2017practical}. Path tracing with extended PPG could substantially reduce the noise of the final image. \copyright 2019 Disney. This image was originally published in  \cite{vorba2019path}.}
    \label{fig:ppg}
    \vspace{-2mm}
\end{figure}
}

Even with advanced denoising and adaptive sampling methods enabled, sometimes the variance remains high for complex scenes, and both of these methods still need a substantial amount of samples to produce noise-free results. They do not directly address the primary source of MC noise from light transport simulations, which limits their capabilities to reduce the variance.

Path guiding is one of the unbiased adaptive variance reduction techniques \cite{vorba2019path}. The key idea is to sample the ray direction cleverly during path construction so that images could be rendered even with a low amount of samples per pixel. As mentioned in Sec.\ref{sec:pg_basic}, path guiding aims to construct more paths with significant radiance contribution to pixel values while minimizing the number of paths that have zero contribution. This is achieved by learning from historical sampled paths and inferring the scene radiance distribution that could be used for importance sampling remaining paths.

\vspace{-2mm}
\subsection{Path guiding using simple fitting}
In the early development of path guiding, people constructed the local sampling distribution based on functional fitting and some simple heuristics. Jensen et al. \shortcite{jensen1995importance} first traced a set of photons from light sources, and then fitted a hemispherical histogram from nearby photon energies at each surface shading point. The key idea was that photons gave a good approximation of the local radiance field, representing the illumination from light sources. The histogram was further converted to a probability distribution for importance sampling new paths. This work was further extended by \cite{steinhurst2006global}, which added product importance sampling with BSDF and by \cite{budge2008caustic}, which was focused on path-guided caustics. Later Hey et al. \shortcite{hey2002importance} replaced histograms with width-varying cones that were centered around photon directions (i.e., the incident direction of light). One problem of such particle-based guiding methods was the non-uniformity of photon distributions. In the worst case, the fitted sampling distribution could be in terrible shape when there was only one nearby photon, causing the entire path guiding algorithm to stuck. To deal with this problem, the photon emission was also guided based on the importance to the camera, resulting in up to 8 times higher photon density in camera visible regions \cite{peter1998importance}. Lafortune et al. \shortcite{lafortune19955d} proposed the first work that subdivided the space of positions and directions into a hierarchical 5D tree that stored incident radiance of samples, and the tree was then used for importance sampling. 

\subsection{Path guiding using learning algorithms}
Fitting sampling distributions locally through simple histograms and cones limited the adapting ability to detailed variations, sometimes leading to poor distributions even if a large number of photons were used. And when photons were highly non-uniformly distributed, these methods could produce bad fitting results such as delta functions in very sparse regions. Vorba et al. \shortcite{vorba2014line} pointed out that reconstructing sampling distributions from photons was eventually a density estimation problem. They adopted the Gaussian Mixture Model (GMM) to represent the radiance field and a progressive online learning method using the Expectation-Maximization (EM) algorithm to infer the parameters of GMM. The results showed a much superior rendering quality using the same number of emitted photons compared to histograms and cones. Moreover, their experiments supported an important claim that path tracing with guiding could reach comparable or even better performance over advanced bidirectional rendering algorithms such as Bidirectional Path Tracing (BDPT) and Vertex Connection and Merging (VCM), while BDPT and VCM were much more difficult to be integrated into the production pipeline. Subsequent work \cite{herholz2016product} extended it to product importance sampling with both radiance and BSDF represented by GMM, and another work \cite{vorba2016adjoint} added Russian Roulette (RR) to path guiding. Later Dahm et al.  \shortcite{dahm2017learning} observed that the same integral equation governed light transport simulations and reinforcement learning. More specifically, the rendering equation (Equ.\ref{eqn:re}) could be written in the form of Q-learning:
\begin{align}
\begin{split}
    Q^{\prime}(p_{x}, \omega_{o}) = &(1-\alpha) \cdot Q(p_{x}, \omega_{o}) +\alpha \cdot[ L_{e}(p_{y}, \omega_{o})\\& + \int_{H}Q(p_{y}, \omega_{i})f(p_{y}, \omega_{o}, \omega_{i})|cos(\theta_{i})|d\omega_{i}]
    \label{eqn:ql}
\end{split}
\end{align}
where $Q$ is the incident radiance, light source radiance $L_{e}$ is regarded as the reward, the current shading position $p_{x}$ represents the current state, and the next shading position $p_{y}$ is the next state if we shoot the ray from $p_{x}$ in direction $\omega_{o}$ (i.e., action). The learning algorithm was trained by updating $Q$ according to Equ.\ref{eqn:ql} and it could find the policy to take the best action so that sampled scattering directions could maximize the total reward (i.e., radiance contribution). 

Muller et al. \shortcite{muller2017practical} proposed a more practical path guiding approach which had been adopted by Disney for rendering Frozen 2 (\copyright 2019 Disney). They combined a hierarchical spatial binary tree and directional quadtree with adaptive binning (SD-tree) to represent the incident radiance field. The historical sample radiance was cached in this hybrid data structure. The progressive reinforcement learning was then applied for iteratively updating the radiance distribution and adaptively adjusting the SD-tree structure. Later Muller et al. \shortcite{vorba2019path} further improved the original method by adding some new modules. First, the original method threw away rendered images of previous iterations. It reset the image to black and restarted the guided path tracing from scratch for every progressive learning iteration. This was because the path-traced image could have noise and fireflies when the online path guiding was not trained for enough iterations to produce good sampling distributions, and these noisy images could pollute final results if not being thrown away. However, a lot of useful samples would be wasted in this way. To resolve this problem, they proposed to weighted combine images of different iterations according to their inverse variances so that we could get noise-free results faster without throwing away previous samples. Second, when caching the radiance, the original method splatted the radiance into the hybrid SD-tree based on the nearest neighbor search, which would sometimes cause visible artifacts. They proposed to filter the discrete spatial and directional tree structure, which removed the artifacts almost entirely. Finally, they incorporated BSDF sampling with guiding by one-sample multiple importance sampling (MIS). They also showed that the optimal MIS weights could be learned by minimizing Kullback-Leibler divergence using gradient descent optimization. Over the past few years, this method has become the most popular path guiding technique, and it has been adopted for movie productions. 

\Comment{
\begin{figure}[t]
    \includegraphics[width=\linewidth]{figs/cloud.png}
    \caption{Rendering volumetric clouds using a deep neural network to predict multi-scattered radiance. The rendered images using the predicted radiance look indistinguishable from brute-force path tracing results. \copyright 2017 Disney. This image was originally published in \cite{kallweit2017deep}.}
    \label{fig:cloud}
    \vspace{-3mm}
\end{figure}

\begin{figure*}[t]
    \includegraphics[width=\textwidth]{figs/art_manip.pdf}
    \caption{Original rendered images are usually further manipulated by artists to meet aesthetic needs. The control over each rendering and compositing process is crucial for art-directed movie production. Moana \copyright 2018 Disney. These images were originally published in \cite{burley2018design}.}
    \label{fig:art_manip}
\end{figure*}
}

Besides the progressive online reinforcement learning, people have also tried path guiding using deep neural networks. Muller et al. \shortcite{muller2018neural} designed a U-shape neural net that could directly generate MC samples and their probability densities. They also demonstrated that neural nets were very good at approximating the true incident radiance field, and the learned distributions were much sharper than the ones learned by GMM and SD-tree. The proposed network warped the uniform distribution to the desired distribution by transforming random variables into the desired samples. However, to use the new samples within the MC context, the probability density must also be evaluated for each sample. To compute such probability, the network worked as a chain of differentiable parametric bijective functions that related the uniform probability density with the warped density by Jacobians. The neural network parameterized each of these bijective functions, and it was trained by minimizing the Kullback Leibler-divergence between the desired distribution and the one represented by the network. They also extended this network to handle BSDF sampling as well to achieve product path guiding holistically, and to optimize one-sample MIS weights for linearly blending between the learned radiance distribution and BSDF distribution. This gave their network full control over the MC importance sampling process. Recently, another work \cite{bako2019offline} was the first to try offline learning for scene-independent path guiding, which avoided the cold start problem in online methods. The key idea was to use the neighboring local samples around each shading point with additional features for reconstructing the radiance field. Using the neighborhood samples and training the network offline allowed the whole system to start guiding subsequent samples as quickly as possible after caching only a small amount of initial samples, which significantly shortened the expensive online training. More specifically, the network learned to regress/interpolate the full sampling distribution using nearby cached sparse samples, since the radiance field was assumed to be coherent within a small local region. A similar neural network \cite{zheng2019learning} was proposed to generate importance samples also by nonlinear warping, but this time in the Primary Sample Space (PSS) that could then be mapped to the path space. All these neural path guiding works have achieved superior performance, especially for scenes with difficult light transport scenarios, and are potentially much faster using GPUs.

In movie production, path guiding techniques are still under continuing development. Studio artists are now trained to place light sources in a way that makes light transport simulation easier. Consequently, guiding is only enabled for some difficult indirect light transport cases in Frozen 2 (\copyright 2019 Disney) and Alita: Battle Angel (\copyright 2018 Twentieth Century Fox). In fact, the gain over vanilla path tracing is marginal for easy lighting conditions.

\vspace{-4mm}
\section{Other special rendering topics}
\label{sec:other}

Although denoising and path guiding are the two most important techniques for production rendering, over the years, people have also developed other learning-based frameworks on specific rendering topics. Kallweit et al. \shortcite{kallweit2017deep} studied high-albedo volumetric cloud rendering problem. The brute-force path tracing was computationally expensive for this case since light could be scattered by hundreds or thousands of times within the volume. Otherwise, the cloud would look darker if not enough scattering events were simulated. Therefore, they proposed a simple MLP network that encoded the cloud densities hierarchically and predicted the radiance contribution of multiple scattering (i.e., indirect scattered illumination). Their network could render clouds about 2000 times faster on a GPU than the brute-force path tracing on a 48-thread CPU.

Most recently, Zhu et al. \shortcite{shilin2020deep} applied the MLP network to the density estimation of photon mapping (PM). As mentioned in Sec.\ref{sec:pm_basic}, PM algorithms were used to create high-quality caustics effects. The problem with traditional PM was that a large number of photons were required for generating converged results. To speed up this algorithm, they designed a deep neural network that predicted a kernel for each shading point to aggregate its neighboring photon energies. The network took nearby photons of each shading point and extracted features out of them, which was followed by kernel prediction and radiance regression. This approach can produce a high-quality reconstruction of caustics using only sparse photons (i.e., an order of magnitude fewer photons).

Other work such as using GAN to generate micro-scratches on surfaces \cite{kuznetsov2019learning}, using CNNs to create better sampling point patterns \cite{leimkuhler2019deep}, and gradient-domain light transport simulation using a multi-branch autoencoder \cite{guo2019gradnet} also demonstrated the effectiveness of neural networks. All of these techniques have the potential to be adopted by future movie productions.

\vspace{-1mm}
\section{Discussion and Conclusion}
\label{sec:conclusion}

In this paper, we surveyed machine learning approaches for rendering high-quality frames in movie production. Among these techniques, denoising and path guiding have already been adopted by multiple studios. However, there are still some open problems. As for denoising, even the most advanced neural networks can fail in some particular cases such as regions of hair, fur, or small bumped shiny surfaces. In fact, improving the generalizability of neural networks is still actively being studied in the machine learning community. As for path guiding, the cold start of online methods and generalization capability of offline methods are problems that could reduce or even erase the gain over vanilla path tracing. In volumetric rendering, light could be scattered thousands of times within the volume. Extending path guiding to every one of these bounces may open up new problems. Besides, machine learning systems are not only good at fitting functions and removing noise, but also good at generating rare artifacts which are even harder to remove altogether. These artifacts look less visible in most interactive gaming applications, but are big problems for film productions that aim for "perfect pixels". Therefore, a single pass of ML inference may not be enough. Combining neural networks with some progressive algorithms (e.g., denoising + adaptive sampling) is one workaround proposed to relieve this problem. Finally, there are trade-offs between neural systems speed and brute-force path tracing speed. If ML takes an even longer time to process the data than shooting a lot more rays/samples in traditional algorithms to produce comparable results, there would be no reason to make the renderer more complicated. Considering that most production renderers are running on CPUs and neural nets are faster on GPUs, the CPU/GPU hybrid rendering pipeline is desirable to avoid the additional data transfer cost.

Besides technical problems, many learning-based rendering algorithms lack sufficient control over the results. In general, artists often want to adjust the level/strength of each process, such as using a knob to control how much denoising should be applied. This problem dramatically defers these methods to be used by the movie industry since the controllability and flexibility are crucial in the production pipeline. Moreover, studio renderers usually have unique design choices (e.g., ray batching and sorting) to handle complex geometries, large texture maps, and difficult light transports. Therefore, techniques that are proposed by research institutions could have very different performance on production renderers and large-scale movie scenes. 

Despite the fact that learning-based rendering methods have some remaining problems, they have demonstrated promising results in lots of rendering sub-fields. Currently, we have not fully understood the principles of machine learning yet, and most people treat it as a black box. However, this should not prevent us from using it in rendering movies and uncovering more insights from our future attempts. The machine learning and rendering systems could be tightly integrated into a fully automated fast rendering pipeline for producing future animated feature films.

\vspace{-3mm}

\bibliographystyle{ACM-Reference-Format}
\bibliography{reference}

\end{document}